# An Artificial Intelligence Browser Architecture (AIBA) For Our Kind and Others:
*A Voice Name System (VNS) Speech implementation with two warrants, Wake Neutrality and Value Preservation of Personally Identifiable Information*

Brian Subirana
Massachusetts Insitute of Technology
Harvard University

**Abstract:** *Conversational commerce, first pioneered by Apple's Siri, is the first of may applications based on always-on artificial intelligence systems that decide on its own when to interact with the environment, potentially collecting 24x7 longitudinal training data that is often Personally Identifiable Information (PII). A large body of scholarly papers, on the order of a million according to a simple Google Scholar search, suggests that the treatment of many health conditions, including COVID-19 and dementia, can be vastly improved by this data if the dataset is large enough as it has happened in other domains (e.g. GPT3). In contrast, current dominant systems are closed garden solutions without wake neutrality and that can't fully exploit the PII data they have because of IRB/Cohues-type constraints.*

*We present a voice browser-and-server architecture that aims to address these two limitations by offering wake neturality and the possibility to handle PII aiming to maximize its value. We have implemented this browser for the collection of speech samples and have successfully demonstrated it can capture over 200.000 samples of COVID-19 coughs. The architecture we propose is designed so it can grow beyond our kind into other domains such as collecting sound samples from vehicles, video images from nature, ingestible robotics, multi-modal signals (EEG, EKG,...), or even interacting with other kinds such as dogs and cats.*

1. **Introduction**

We have published elsewhere the need to design a technical architecture that can support an ecosystem of artificial intelligence players while optimizing value for those involved, which we understand means juggling many different design priorities:
- **A Voice Name System to name any object in the world:** Multiple devices may be present and multiple end-points may exist giving rise to conflicts that need to be resolved [2, 6].

- **A Wake Neutrality architecture:** We would like the infrastructure to provide equal access to all players such as the phone system does once you know a phone number, a property we call Wake Neutrality [13] for its similarity in spirit to Net Neutrality.
- **A Set of Common Biomarkers:** We would like the infrastructure to develop AI models that have some commonality across them [4, 5].
- **A Standardized Brain Model such as MIT's CBMM Brain Model:** For our kind [2] we would like to have a reference brain model so that we don't have to re-invent everything from scratch every time. For other kinds we'll need similar reference models such as acustic markers for automobile diagnosis.
- **A Common Set of Use Cases such as those suggested by MIT Open Voice:** By developping common use cases, people can grow services around them [15].
- **A Common Language such as Huey:** Interactions should have as much standardization as possible so that creators and consumers don't have to re-learn similar concepts [3, 15].
- **Application to other Kinds:** Conversational commerce will be everywhere and cars, and even bricks will have their own personalities that will need to be supported [11].
- **Legal Programming:** In an Internet of Things world, there are many hurdles to legal compliance that need to be addressed, especially PII and privacy in general [1, 6, 11, 12]
- **Common Architectural Boundaries:** To support the development of hardware and software products, clear boundaries need to be established [3, 7, 8, 10, 13].

In the rest of this paper we provide the specs for a browser-and-server architecture that was the basis for an implemented system that we used in my laboratory to collect speech samples for COVID-19 research [4, 5]. I hope it will inspire others to create similar versions that can be open sourced and shared, and that compatibly grow the architecture into other applications, modalities and kinds.

## 2. AIBA 1.0: Highlights

**Summary of functionality:** The user opens the app and sees a record button with (or without) instructions on what to record. The user can start recording by pressing the microphone icon as soon as the app is opened, except the first time that the terms are to be shown and accepted. The app records all recordings with a unique *hash_number*, always the same for every phone, that the user never sees. Everything else is optional and happens in the configuration options (introduce personal non-identifiable information, select language) or in the server (set instructions of what to record).

**Highlights of the suggested architecture:** The server sends a number of instructions on what to record to the App (they are included in the <app_runtime_config_file>). The App runs the user through them. The App sends the server the voice samples. In addition, if the user makes changes to the configuration, these are sent to the server. There are some exceptions, as when no instructions are given (free recording), or as when the server processes the audio and sends some feedback on whether the user has tested positive or not (the processing option will be introduced after the models have been built), behavior which will be defined by an <Engine_number>. Tests will be recorded in the subdirectory of the server app directory (name of the subdirectory is the *hash_number* of the phone).

**Note on the tested version:** We have developed two versions of the app. The MIT Voice App was written in React Native and there is also a Web version of the app that was tested without interruption for two years at http://opensigma.mit.edu. It was used to collect over 200.000 samples of coughs during the first months of the COVID-19 pandemic.

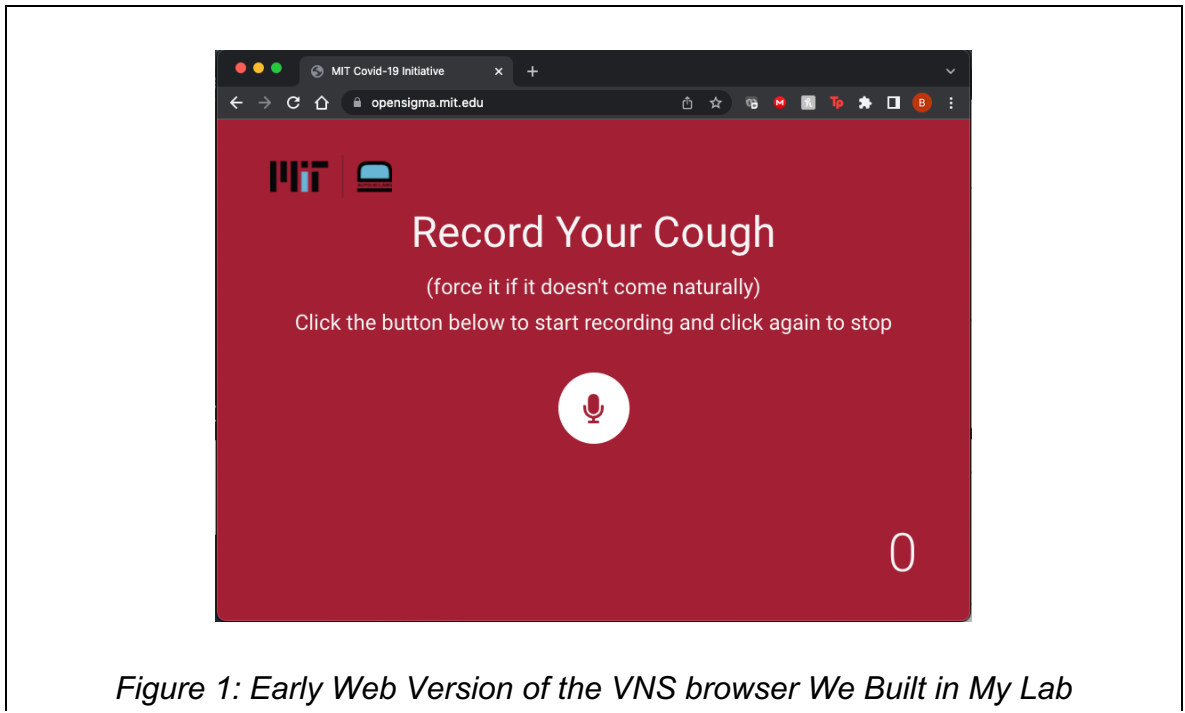

*Figure 1: Early Web Version of the VNS browser We Built in My Lab*

## 3. AIBA 1.0: Basic interface

AIBA 1.0 is centered on collecting speech and natural language input from our kind:

- Interface design
  - Screen opens and terms are accepted (this screen is only shown the first time)
  - Second screen (loads default app language and default words in the app from the server-side <app_runtime_config_file>, if there is no internet connection or it's slow)
    - Text output: <text_request> (e.g. "Digues els números del 0 al 10")
    - Record button
    - Text input: <text_input_box> (e.g. "Diguens com et trobes").
    - Settings button
    - A number (corresponding to the number of recordings, which keeps increasing)
    - Behavior of this screen:
      - When pressing the record button some feedback is given so that the user knows recording is on. Recording happens while this feedback is given and for as long as the number of the pair given in the pair. Recording can be stopped by pressing the record button again."
      - Settings sends you to settings
      - The number increases after each recording
      - The text input is a box where to write text associated with the sample (the user may type whether it has fever or any comorbid conditions)
      - If the **<app_runtime_config_file>** is empty, the user keeps recording what it wants (it's a free recording mode for the app without specific instructions). Otherwise, it round robins around the options.
  - Third screen –
    - if the **<app_runtime_config_file>** ends with a text without seconds then this text is displayed (e.g. "thank you please record again tomorrow"), and no other options are given to the user but a button that says: "start over". When the user presses it, we go back to screen two.

- If the directory <phone_hash> has a text file and/or audio file, the text is displayed and/or the audio played. E.g., "This text could be generated from the response of the <Engine_number>" (e.g. results from COVID-19 test). If there is no response from the server we keep recording
- In general, if there is no connection with the server we should be collecting samples (and send them when there is internet coverage). This may happen in the basement of a Hospital.

- Settings option
  - Choose language. Selection is added to the <local_config_status_file>
  - Introduce password and "codi d'estudi".
  - Tell us about you:
    - Opens up a menu with field texts associated to the requests of file <personal_information_request>. For the rapid implementation we can give a fixed set of questions if this is too complicated. We need a fast turnaround.
  - Generate single numbers to be shared with neighbors (i.e. different numbers to give to people related to you). You can press it indefinitely. These numbers are also added to the <local_config_status_file>
  - Introduce neighbors' numbers (e.g. different numbers that others give you). Numbers added to the <local_config_status_file>
  - See terms
  - Run_time_file_config_number. Number added to the <local_config_status_file>
  - Reset number of recordings. Resets number to zero. Two numbers are updated. In the <local_config_status_file>. <Total_count>=<Total_count>+<Current_count> and after <Current_count>=0. and the total count
  - Fix color
  - <reset_time> it's the time before a session becomes inactive and a reset is established.
  - <Engine_number> identifies the engine to provide the answer as to whether the user has the virus or not according to the algorithm
  - <VNS_number> describes the server that will interact with the phone (it can be an ip address or just a name).
  - <Dynamic_VNS_toggle> the default is "on" but it can be set to "off" by the user

o  <Dynamic_VNS> the server that will be used if the above is on. As a default if none is established, the app should point to http://voice.mit.edu

**3. Configurable options from the server-side:**

Any element of any of the <app_runtime_config_files> can be changed / added in real-time. E.g., a new file can be introduced to reflect a new collection strategy for another COVID-19 study, as well as a default <Engine_number>.

**Other considerations:**

- A back-end server is required to collect the samples
- All the recordings from the same phone should have a unique hash that is invisible to the user (stored in the phone but the user can not see it). Samples should also be saved with an associated timestamp.
- It's a must to have multiple languages. In our implementation we had English, Spanish and Catalan to start
- Source code will be made available so that it can be updated and improved by others
- Server will provide a daily download of the audio files with the associated metadata.

**4. Files involved:**

**<app_runtime_config_file>**: ← it's a JSON file to be downloaded

- This csv file contains a <string> and 0 to four lists with the basic recordings to be done by the user. Each list has pairs <"text", seconds>, associated with a recording. E.g. <"tossi 10 segons", 10> or <"digui ommm amb la m prolongada 10 segons si pots", 12>.
    o List selection. If there is only one list, that one is used. Otherwise, a pop up screen with the <string> is displayed so that the user can select one of the list. This means that two, three or four buttons are displayed, one per list, to select which list is the one to be used.
    o Special situation: If no time in one of the pairs, the recording should be crossed and the user must input a text and press a new button that says <no_recording_text> (e.g. "Recording de-activated, submit text only"). This button only appears in this situation.

- The name of this file should be name_number.csv where name is "app_runtime_config_file" and number can be changed. The phone will request the app_runtime_config_file with a number and that is the file to be sent.
- Special situations are:
    - "empty file": app keeps recording without questions
    - File starts with <no_recording, 0> then recording option is deactivated and the user can simply input text.

**<sample_server_upload_file>** ← it's a JSON file to be uploaded. This file contains all options of the setting file and is only uploaded if:

- The file has been changed by the user
- More than <reset_time> minutes have occurred since the last recording

**<personal_information_request>** ← In the future, this file will contain the questions that can be asked. We can not ask for personal information that may identify the person. For the first version the questions can be fixed. This means we won't ask for a name, or day and month of birth (year up to 89 or over 90 is ok). We imagine three types of questions:

- Text: E.g., tell us about how you are feeling
- Pull down: E.g. year of birth
- Multiple-Choice: E.g. have you been diagnosed with COVID-19?

Detailed questions could be asked in the first implementation of the COVID-19 study:
- Country
- ZIP
- Age (1-89 and 90 or more - we cannot request age above 90)
- Have you been diagnosed with COVID-19?
    - if so, when?
- Do you have any of the following:
    - Fever
        - If so, how much?
    - Did you cough today?
        - If so, how much
- Tell us about any other symptoms that you think are relevant

## 5. Details of some of the Variables involved:

**< Run_time_file_config_number>** : Identifies the number of the <app_runtime_config_file@number> that is to be associated with this session. For example, in the current scenario it would be a configuration file associated with COVID-19. In the future it may be one associated with another pandemic.

**<max_recording_time>** : maximum recording time (in case no default is set in the particular recording)

**<Engine_number>** This variable is associated with a routine for how the recorded data should be processed, either by sending it to an API endpoint and returning a written response (always associated with <phone_hash>), or otherwise internally processing the sample.

**Old Screens Shown as Examples:**

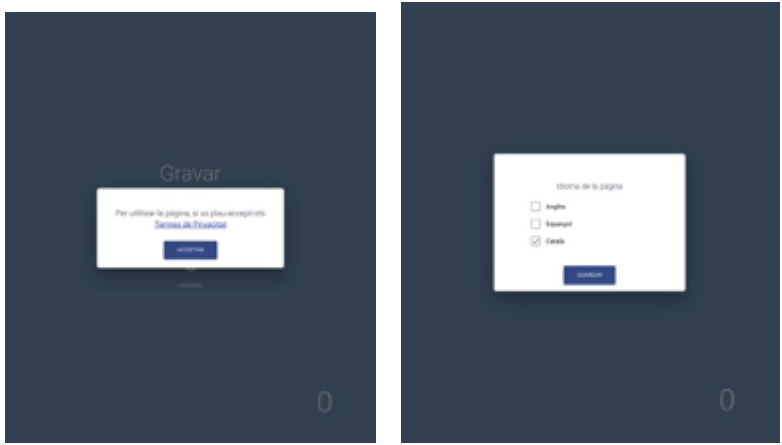

Figure 2:
*Left:* Screen with the terms.
*Right:* Screen with a pop-up to select what to record.

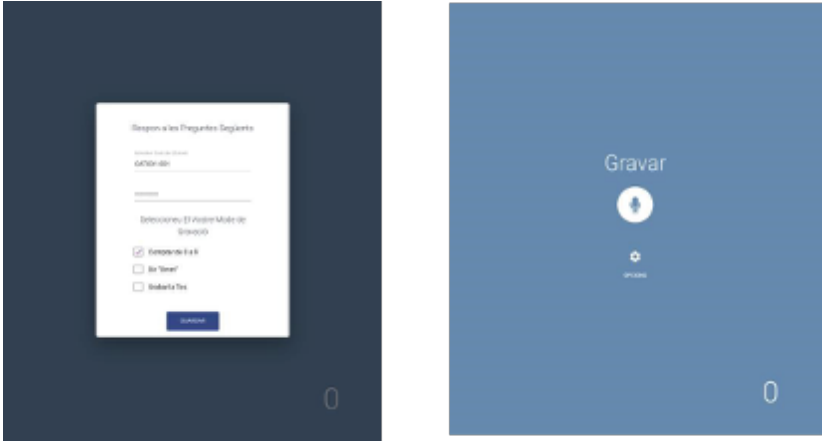

Figure 3:
<u>Left:</u> Screen with settings that was amplified in the final version.
<u>Right:</u> Screen with recording. Text box was added for particular tasks.

## 6. Conclusion

We have outlined the motivation to design a new architecture for artificial intelligence transactions and referenced the research that lead up to this specifications. There are some efforts, such as the Open Voice Network (OVON), that have similar goals to the ones of the effort reported here. The Open Voice Network is a neutral, non-profit industry association operating as a directed fund of The Linux Foundation whose origins can be traced to research we did in my laboratory, the MIT Auto-ID Laboratory as shown in their About page of http://www.openvoicenetwork.org (Accessed March 2022).

The aims of AIBA are very much in line with those of OVON as shown in their vision and mission statement. However, our technical progress provides evidence that perhaps we these statements are too narrow for what the world needs, which is broader application, multi-modal support and extensions beyond our kind into machines, nature and the broad internet of things infrastructure.

**Bibliography**

[1] Malcolm Bain and Brian Subirana. Legal programming Vol 49, No. 9, pp.57-62. Communications of the ACM. September 2006. (Co-authored with M. Bain.)

[2] Harris, M., 1989. *Our kind: Who we are, where we came from, where we are going*. Harper & Row Publishers.

[3] Levinson, Harry. 2021. Huey: Intelligent Agents for Natural Human to Machine Communication. Master's thesis, Harvard University

[4] Jordi Laguarta and Brian Subirana. Longitudinal speech biomarkers for automated alzheimer's detection. Frontiers in Computer Science, 3:18, 2021. ISSN 2624-9898. URL https://www.frontiersin.org/article/10.3389/fcomp.2021.624694.

[5] Jordi Laguarta, Ferran Hueto, and Brian Subirana. Covid-19 artificial intelligence diagnosis using only cough recordings. IEEE Open Journal of Engineering in Medicine and Biology, 1:275–281, 2020.

[6] Brian Subirana. Call for a wake standard for artificial intelligence. Communications of the ACM, 63(7):32–35, 2020.

[7] Brian Subirana. "Back to the Future. Anticipating Regulatory Hurdles within IoT Pelotons." The Internet of Things (Chapter 21). Book edited by The American Bar Association 2019, ISBN: 978-1-64105-363-1.

[8] Brian Subirana, Sanjay Sarma, Richard Cantwell, Jon Stine, Mark Taylor, Kees Jacobs, Shannon Warner, Gwendolyn Graman, and Chris Hunt. Time to talk: The future for brands in conversational. capgemini, intel, 2017.

[9] Brian Subirana, Alex Gaidis, and Sanjay Sarma. Conversational computer programming: Towards a software engineering research agenda. Unpublished manuscript, 2018.

[10] Brian Subirana, Alexander Gaidis, Sanjay Sarma, Richard Cantwell, Jon Stine, Peter Oliveira-Soens, Esteve Tarragó, Ferran Hueto-Puig, Prithvi Rajasekaran, and Alex Armengol-Urpí. A secure voice name system to wake smart devices. MIT Stephen A. Schwarzman College of Computing Research and Computing Launch Postdoc and Student Poster Session, 2019. URL 10.13140/RG.2.2.12884.65921.

[11] Subirana, B., Haghighi, N., Cantwell, R., & Sarma, S. (2018). Conversational Bricks and The Future of Architecture: Will< Stores> Survive as the Epicenter for< Retail> Activity in Society?. International Journal of Structural and Civil Engineering Research Vol. 7, No. 3, Pages 238-245. doi: 10.18178/ijscer.7.3.238-245

[12] Brian Subirana and Malcolmn Bain. *Legal programming. Designing legally compliant RFID and software agent architectures for retail processes and beyond*. Research monograph published by Ed. Springer, 2005, 313 pages. ISBN: 0387234144.

[13] Subirana, Brian, Renwick Bivings, Sanjay Sarma "Wake neutrality of artificial intelligence devices." In *Algorithms and Law*, eds. M. Ebers and S. Navas. *Cambridge Univ. Press*, pp. 235-268. (2020)

[14] Brian Subirana, Ferran Hueto, Prithvi Rajasekaran, Jordi Laguarta, Susana Puig, Josep Malvehy, Oriol Mitja, Antoni Trilla, Carlos Iv́an Moreno, José Francisco Muñoz Valle, et al. *Hi sigma, do i have the coronavirus?: Call for a new artificial intelligence approach to support health care professionals dealing with the covid -19 pandemic*. arXiv preprint arXiv:2004.06510, 2020.

[15] Brian Subirana, Harry Levinson, Ferran Hueto, Prithvi Rajasekaran, Alexander Gaidis, Esteve Tarragó and Peter Oliveira-Soens. *The MIT Voice Name System*. White Paper. arXiv preprint 2022.